\documentclass[12pt,a4paper]{article}
\usepackage{setspace}
\usepackage{graphicx}
\usepackage{multirow}
\usepackage{amsmath,amssymb,amsfonts}
\usepackage{amsthm}
\usepackage{mathrsfs}
\usepackage[title]{appendix}
\usepackage{xcolor}
\usepackage{textcomp}
\usepackage{manyfoot}
\usepackage{booktabs}
\usepackage{algorithm}
\usepackage{algorithmicx}
\usepackage{algpseudocode}
\usepackage{listings}
\usepackage[numbers]{natbib}
\usepackage[left=2.54cm, right=2.54cm, top=2.54cm, bottom=2.54cm]{geometry}
\allowdisplaybreaks[3]
\title{Domain Wall Bianchi Type $VI_0$ Universe in $f(R,T)$ Gravity}
\author{S. P. Hatkar$^1$, D. P. Tadas$^{2}$\footnote{Corresponding author E-mail: dtadas144@rediffmail.com}, S. D. Katore$^3$ \\
	$^1$Department of Mathematics, A.E.S. Arts, Commerce and Science\\ College, Hingoli-431513, India.\\
	$^{2*}$Department of Mathematics, Toshniwal Arts, Commerce and Science College,\\ Sengaon-431542, India.\\
	$^3$Department of Mathematics, Sant Gadge Baba Amravati University\\ Amravati-444602, India.}
\date{}
\begin{document}
\maketitle
\onehalfspacing
\begin{abstract}
	We consider the Bianchi type-$VI_0$ space time with domain walls in the framework of the modified $f (R, T)$ theory of gravitation. To solve the field equations, we assume that the shear scalar $(\sigma)$ is proportional to the expansion scalar $(\theta)$. We also consider the parametrization of the equation of the state parameter of barotropic fluid and discuss the effect on domain walls. It has been observed that the domain wall may behave like dark energy. Some physical parameters are also discussed in detail.
\end{abstract}
 \section{Introduction}
	The first few minutes after the Big Bang are important in the history of the universe. When the temperature drops to some critical point, topological defects occur due to the phase transition of the universe. The idea of spontaneous symmetry breaking prompts that the topology of the vacuum manifold $M$ is identified as domain walls. Topological defects are remnants of the phase transitions that may have occurred in the early universe \cite{Vilenkin 1985,Rajaraman 1987,Yilmaz 2006}. Hill et al. \cite{Hill et al. 1989} have proposed that domain walls are important in the formation of galaxies. Recently, Reddy and Naidu \cite{Reddy and Naidu 2007} have analysed thick domain walls in scale-covariant theory of gravitation. Lazanu et al. \cite{Lazanu et al. 2015} have studied the contribution of domain walls to the cosmic microwave background power spectrum. Adhav et al. \cite{Adhav et al. 2007} have investigated thick domain walls in the Brans-Dicke theory of gravitation.

	Einstein's general theory of relativity (GTR) was first proposed in 1905, and it was later modified by replacing the term $ R $ with $ f(R) $, where $ R $ represents the Ricci scalar, in the Einstein-Hilbert action. The Einstein GTR fails to explain the late time-accelerated phenomena of the universe; hence, modifications have been incorporated into the GTR. It explains the late-time acceleration of the universe. In $f(R)$ theory, no extra arbitrary degree of freedom is added. It is shown that $f(R)$ models are equivalent to Brans-Dicke theory with coupling parameter $\omega=0$ \cite{Chiba 2003}. However, exactly how $f(R)$ infallibly triggers an accelerated expansion of the universe is as yet unexplained. Also, the $f(R)$ theory has stability and viability issues \cite{Carroll et al. 2004,Nojiri et al. 2003,Capozziello et al. 2003}. In $f(R)$ theory, Amendola et al. \cite{Amendola et al. 2007} investigated the modification of the conventional matter-dominated epoch for dark energy scenarios. Nojiri and Odintsov \cite{Nojiri and Odintsov 2011} propose a classical extension of general relativity as a gravitational alternative to early-time inflation and cosmic acceleration. They investigate modified theories that are focused on local experiments, such as $f(R)$ gravity, non-local gravity, power-counting renormalizable covariant gravity, string and scalar-tensor theory. Nojiri et al. \cite{Nojiri et al. 2017} presented a brief overview of modified gravity theories in cosmology, focusing on inflation, bouncing cosmology, and the late-time acceleration  phase. They investigate the qualitative aspects of the dark energy period, concluding that these theories may explain our universe and provide a plausible inflationary epoch.
	
	Recently, Harko et al. \cite{Harko et al. 2011} proposed a more general model of modified theory known as the $f(R,T)$ theory of gravity, in which $f(R,T)$ is an arbitrary function of the Ricci Scalar $R$ and the trace $T$ of the energy momentum tensor. The coupling between matter and geometry leads to extra acceleration in $f(R,T)$ theory. There is a choice of the functions $f(R,T)$ depending on the nature of the matter field \cite{Sahoo et al. 2016}. Harko et al. \cite{Harko et al. 2011} pointed out that the $f(R,T)$ gravity model depends on the source term, which represents the variation of the matter stress energy tensor with respect to the metric. The simplest model, $f(R,T)=R+\lambda T$, is equivalent to a cosmological model with an effective cosmological constant. The term $f(T)$ in gravitational action modifies the gravitational interaction between matter and curvature. In $f(R,T)$ theory, cosmic acceleration comes not only from geometrical contribution but also from matter content \cite{Singh and Singh 2014}. The $f(R,T)=\mu R+\gamma T$ model shows quintessence-like behaviour of the Eos parameter \cite{Pasqua et al. 2013}. Reddy et al. \cite{Reddy et al. 2012} investigated the Bianchi type-III space time in the $ f(R,T) $ gravity and showed that the model has no initial singularity and late-time rapid expansion for large $t$. Rao and Neelima \cite{Rao and Neelima 2013} investigated Bianchi type$-VI_{0} $ cosmological models and discovered that $ f(R,T) $ affects the matter distribution but has no effect on space-time geometry. Singh and Singh \cite{Singh and Singh 2015} discussed the behaviour of a scalar field in $f(R,T)$ gravity theory within the context of a flat FRW cosmological model that describes the early and late time evolution of the universe. Mishra et al. \cite{Mishra et al. 2016} investigated Bianchi type VIh space time in the presence of perfect fluid and discussed the dynamical feature of the models in $f(R,T)$ gravity. Recently, Chaubey and Shukla \cite{Chaubey and Shukla 2017} have obtained the different cosmological solutions of Bianchi models in the framework of $f(R,T)$ gravity.

	Recently, studies of the domain wall have attracted a lot of researchers, and these studies of the domain wall seem to focus on using different theories of gravitation. Sahoo and Mishra \cite{Sahoo and Mishra 2013a} have studied string and domain walls for plane symmetric space times. Mahanta and Biswal \cite{Mahanta and Biswal 2012} have studied string clouds and domain walls with quark matter in Lyra geometry. Maurya et al. \cite{Maurya et al. 2020} investigated domain walls and quark matter in Bianchi type-V universe with observational constraints in $ f(R,T) $ gravity. Further, the tension of domain walls plays an important role in the study of cosmological models of domain walls. The effect of tension on domain walls in higher-dimensional space times is also investigated by Rahaman and Kalam \cite{Rahaman and Kalam 2002}. Biswal et al. \cite{Biswal et al. 2015} have studied Kaluza-Klein domain walls in $f(R,T)$ gravity. Negative tension of domain walls shows that they are invisible, which means that there is energy transfer from domain walls to other forms of matter \cite{Khadekar et al. 2009}. Some authors \cite{Pradhan et al. 2007,Tiwari et al. 2017} have investigated the domain walls cosmological models in different theories.

	With the motivation of the above discussion, we have studied Bianchi type-$VI_{0}$ domain wall cosmological models in the framework of the $f(R,T)$ theory of gravity. Section 2 is devoted to metric and $f(R,T)$ gravity. The field equations for the Bianchi type-$VI_{0}$ space time in the presence of a domain wall are presented in section 3. The solutions of the field equations determined in section 4 for the case  $C_1\neq0$. In section 5, the solution of the field equation is obtained for the case $C_1=0$. The conclusion is presented in section 6.

	\section{Metric and $f(R,T)$ gravity}
	The modified theories of gravitation are aimed to address the problem of gravitational interaction. Renormalizable theories of gravitation are based on the inclusion of higher-order terms of curvature invariant in the Lagrangian. The action for $f(R,T)$ gravity theory is given as
 \begin{equation} \label{Eq:1} 
 	S=\int \sqrt{-g} \left(f(R,T)+L_{m} \right)\,  d^{4} x, 
 \end{equation}
 	where $L_{m}$ is the Lagrangian density of the matter. The field equations of the $f(R,T)$ theory of gravity are obtained by varying action $S$ with respect to $g_{ij}$, which is given by
 \begin{equation} \label{Eq:2}
 	\begin{split}
 		f_R\left(R,T \right)R_{ij}-\frac{1}{2}f(R,T)g_{ij}+\left(g_{ij}\Box-\bigtriangledown_i\bigtriangledown_j \right)f_R\left( R,T\right)\\=8\pi T_{ij}-f_T\left(R,T \right)T_{ij}-f_T\left(R,T \right)\Theta_{ij}
 	\end{split}
 \end{equation}
 	where $T_{ij}=-\frac{2}{\sqrt{-g}}\frac{\delta\sqrt{-g}}{\delta g_{ij}}L_m,$ $\Theta _{ij} =-2T_{ij} +g_{ij} L_{m} -2g^{\alpha \beta } \frac{\partial ^{2} L_{m} }{\delta g^{ij} \partial g^{\alpha \beta }}$, $f_{R} =\frac{\partial {\rm \; }f\left(R,T\right)}{\partial R} $ and $f_{T} =\frac{\partial {\rm \; }f\left(R,T\right)}{\partial T} $.\\
 	We assume the function $f(R,T)$ in the following form:
 \begin{equation} \label{Eq:3}
 	f(R,T)=R+\mu T 
 \end{equation}
 	The late-time dominance of the domain wall network over the energy density provides a very small contribution to anomalies in the cosmic microwave background (CMB). The frozen domain wall is a proposed candidate of dark energy \cite{Avelino and Sousa 2015}. The energy momentum tensor for a thick domain wall of the following form \cite{Sahoo and Mishra 2013b} is given by
 \begin{equation} \label{Eq:4}
 	T_{ij}= \rho\left(g_{ij}+u_{i}u_{j} \right)+pu_{i}u_{j}
 \end{equation}
 	where $\rho$ and $p$ stand for energy density and pressure of the domain wall, respectively, and $u_{i}$ is the unit space-like vector in the same direction with $u_{i}u^{i}=-1$. Here $\rho=\rho_b+\sigma_d$ and $p=p_{b}-\sigma_d$, in which the quantities $\rho_b$ and $p_b$ stand for energy density and pressure of the barotropic fluid, and $\sigma_d$ is the tension of the domain wall. It is believed that there is an anisotropic phase of evolution in the universe, i.e., the universe may be anisotropic in the early phase of evolution. Bianchi type $I-XI$ space times are important in the study of the anisotropic universe. We consider the Bianchi type $VI_0$ space time as
 \begin{equation} \label{Eq:5}
 	ds^2=-dt^2+A^2dx^2+B^2e^{-2\alpha x}dy^2+C^2e^{2\alpha x}dz^2
 \end{equation}
 	where $A,$ $B,$ $C$ are the functions of $t$ and $\alpha$ is a non-zero constant.
 	
 	Bianchi type $VI_0$ space time is of particular interest since it is a simple generalization of Bianchi type space and, at the same time, sufficiently complex. Primordial helium abundance and isotropization in a special sense are explained by Barrow \cite{Barrow 1984} with the help of Bianchi type $VI_0$ space time. Weaver \cite{Weaver 2000} used magnetism to investigate the Bianchi type $VI_0$ metric. Recently, the Bianchi type $ VI_{0} $ space time has been studied by Pradhan and Bali \cite{Pradhan and Bali 2008} in general relativity, whereas Rao and Neelima \cite{Rao and Neelima 2013} and Ram et al. \cite{Ram et al. 2013} in $f(R,T)$ theory of gravity, using different physical fluids.
 
\section{Field equations}
	In the $f(R,T)$ theory, the coupling of matter and gravity contributes equally to cosmic acceleration. The field equations of $f(R,T)$ gravity are derived in metric formalism as 
\begin{equation} \label{Eq:6}
	\frac{\ddot{B}}{B}+\frac{\ddot{C}}{C}+\frac{\dot{B}\dot{C}}{BC}+\frac{\alpha^2}{A^2}=\left(8\pi+5\mu \right)\rho+\mu p 
\end{equation}
\begin{equation} \label{Eq:7}
	\frac{\ddot{A}}{A}+\frac{\ddot{C}}{C}+\frac{\dot{A}\dot{C}}{AC}-\frac{\alpha^2}{A^2}=\left(8\pi+5\mu \right)\rho+\mu p
\end{equation}
\begin{equation} \label{Eq:8}
	\frac{\ddot{A}}{A}+	\frac{\ddot{B}}{B}+\frac{\dot{B}\dot{A}}{AB}-\frac{\alpha^2}{A^2}=\left(8\pi+5\mu \right)\rho+\mu p
\end{equation}
\begin{equation} \label{Eq:9}
	\frac{\dot{A}\dot{B}}{AB}+	\frac{\dot{A}\dot{C}}{AC}+\frac{\dot{B}\dot{C}}{BC}-\frac{\alpha^2}{A^2}= -\left(8\pi+\mu \right)p+3\mu \rho 
\end{equation}
\begin{equation} \label{Eq:10}
	\frac{\dot{C}}{C}-\frac{\dot{B}}{B}=0
\end{equation}
	where the overhead dot represents the derivative with respect to cosmic time. From equation \eqref{Eq:10}, we obtain
\begin{equation} \label{Eq:11}
	C=\beta B
\end{equation}
	where $\beta$ is an integrating constant. The expansion scalar $(\theta)$ and shear scalar $(\sigma)$ are defined as
\begin{equation} \label{Eq:12}
	\theta = 3H =\frac{\dot{A}}{A}+\frac{\dot{B}}{B}+\frac{\dot{C}}{C}= \frac{\dot{A}}{A}+2\frac{\dot{B}}{B}
\end{equation}
\begin{equation} \label{Eq:13}
	\sigma^2= \frac{1}{2}\left[\sum_{i=1}^{3}H_{i}^2-\frac{1}{3}\theta^2\right]= \frac{1}{3}\left(\frac{\dot{A}^2}{A^2}+\frac{\dot{B}^2}{B^2}+\frac{\dot{C}^2}{C^2}-\frac{\dot{A}\dot{B}}{AB}-	\frac{\dot{A}\dot{C}}{AC}-\frac{\dot{B}\dot{C}}{BC}\right) 
\end{equation}
	where $H= \frac{1}{3}\left(\frac{\dot{A}}{A}+\frac{\dot{B}}{B}+\frac{\dot{C}}{C}\right)$ is the Hubble parameter, and $H_{i}, i = 1, 2, 3$ are the directional Hubble parameters in the directions of the $x, y, z$ axes, respectively. As discussed above, the universe is expanding and accelerating. The expansion scalar is used to predict how fast the universe is expanding mathematically. The shear scalar is the rate of shear of the fluid congruence. According to the theorem, "For any space time in which the matter content consists of a perfect fluid and whose flow vector field forms an expanding geodesic and hypersurface orthogonal congruence, if the cosmological constant is zero, $\rho \propto \theta^2$ implies $\sigma^2 \propto \theta^2$ and $R \propto \theta^2$" \cite{Collins 1977}.\\
	According to recent observations of the velocity-redshift relationship for extragalactic sources, the Hubble expansion of the universe is now isotropic to within $\approx 30$ percent. The red shift studies suggest that $\frac{\sigma}{H} \leq0.30$, where $\sigma$ is shear scalar and $H$ is Hubble parameter \cite{Kristian and Sachs 1966,Thorne 1967}. The importance of this physical condition is discussed by Collins and Ellis \cite{Collins and Ellis 1979}. Using equation \eqref{Eq:11}, the system of field equations \eqref{Eq:6} to \eqref{Eq:10} reduces to three equations in four unknowns $A$, $B$, $p$ and $\rho$. We require one more condition to solve the system. It should be noted that we can assume a physical relationship or a mathematical condition. However, assuming a mathematical condition may lead to an unphysical solution, and assuming a physical condition may lead to an unsolvable equation. Therefore, we assume the physical condition that the expansion scalar ($\theta$) is proportional to the shear scalar ($\sigma$), i.e., $\theta \propto \sigma$. Using equations \eqref{Eq:11}, \eqref{Eq:12} and \eqref{Eq:13}, we get
\begin{equation}\label{Eq:14}
	\frac{1}{\sqrt{3}}\left(\frac{\dot{A}}{A}-\frac{\dot{B}}{B}\right)=l\left(\frac{\dot{A}}{A}-\frac{2\dot{B}}{B}\right)
\end{equation}
	where $l$ is a constant. Equation \eqref{Eq:14} further reduces to
\begin{equation}\label{Eq:15}
	\frac{\dot{A}}{A}=n\frac{\dot{B}}{B}
\end{equation}
	where $ n= \frac{1+2l\sqrt{3}}{1-l\sqrt{3}} $. After integrating, equation \eqref{Eq:15} leads to
\begin{equation}\label{Eq:16}
	A= k B^n
\end{equation}
	where $ k $ is a constant of integration. From equations \eqref{Eq:6}, \eqref{Eq:7}, \eqref{Eq:8} and \eqref{Eq:16}, we get
\begin{equation} \label{Eq:17}
	2\ddot{B}+2(n+1)\frac{\dot{B}^2}{B}=\frac{4\alpha^2}{(n-1)k^2B^{2n-1}}
\end{equation}
	Let $\dot{B}=L(B),$ then $\ddot{B}=L\acute{L}$ with $ \acute{L}=\frac{dL}{dB}$ and we get the reduced equation \eqref{Eq:17} in the following form
\begin{equation} \label{Eq:18}
	L^2B^{2(n+1)}=\frac{\alpha^2}{(n-1)k^2}B^4+C_1^2
\end{equation}
	where $C_1^2$ is the integration constant. We have two possibilities, depending on whether the value of the integration constant is zero or not. Note that the $f(R,T)$ gravity models depend on a source term, so each choice of $L_{m}$ would generate a specific set of field equations. Particular cases of $f(R,T)$ functions, including $f(R,T)= R+\mu T$, are discussed by Harko et al. \cite{Harko et al. 2011}. The importance of this model is that the coupling between matter and geometry becomes effective and time-dependent. We would like to mention that one may get different solutions by taking different values of constants and parameters. Consideration of all possibilities of $f(R,T)$ functions and other parameters is left as work for the future. Therefore, we consider the following two cases:

\section{Case I}
	In this case, we have assumed that the non-zero integration constant i.e. $C_1\neq0$. The equation \eqref{Eq:18} is reduced to following equation:
\begin{equation} \label{Eq:19}
	\frac{B^{n+1}dB}{\sqrt{\frac{\alpha^2}{(n-1)k^2}B^4+C_1^2} }= dt
\end{equation}
	For simplicity, we take a particular value of constant, $ n=0 $ and by solving equation \eqref{Eq:19}, we arrive at,
\begin{equation} \label{Eq:20}
	\frac{du}{\sqrt{C_1^2-\frac{\alpha^2}{k^2}u^2}}= 2dt
\end{equation}
	where $ u=B^2 $.\\
	Simplifying and integrating, we get
\begin{equation} \label{Eq:21}
	B=\sqrt{\left(\frac{kC_1}{\alpha}\right) sin(\frac{2kt}{\alpha})}
\end{equation}
\begin{equation} \label{Eq:22}
	C= \beta \sqrt{\left(\frac{kC_1}{\alpha}\right) sin(\frac{2kt}{\alpha})}
\end{equation}
\begin{equation} \label{Eq:23}
	A= k
\end{equation}
	The scale factors are oscillatory. The oscillatory models are physically important since they are alternatives to inflation. From equations \eqref{Eq:6}-\eqref{Eq:9}, we obtain the expressions for energy density and pressure are as follows:
\begin{equation} \label{Eq:24}
	\rho= \frac{ 6\pi b^2cos^2(\frac{2kt}{\alpha})}{l_{1}sin^2(\frac{2kt}{\alpha})}-{\frac{8\pi \alpha^2 }{3l_{1}k^2}}
\end{equation}
\begin{equation} \label{Eq:25}
	p= \frac{ (-2\pi+\mu) b^2 cos^2(\frac{2kt}{\alpha})}{l_{1}sin^2(\frac{2kt}{\alpha})} +{\frac{ (2\pi+\mu) \alpha^2}{l_{1}k^2}}
\end{equation}
	where $ b=\frac{2k}{\alpha}$ and $l_{1}=64\pi^2+48\pi\mu+8\mu^2$ .

\begin{figure}[htb]
	\centering
	\includegraphics[width=0.7\linewidth]{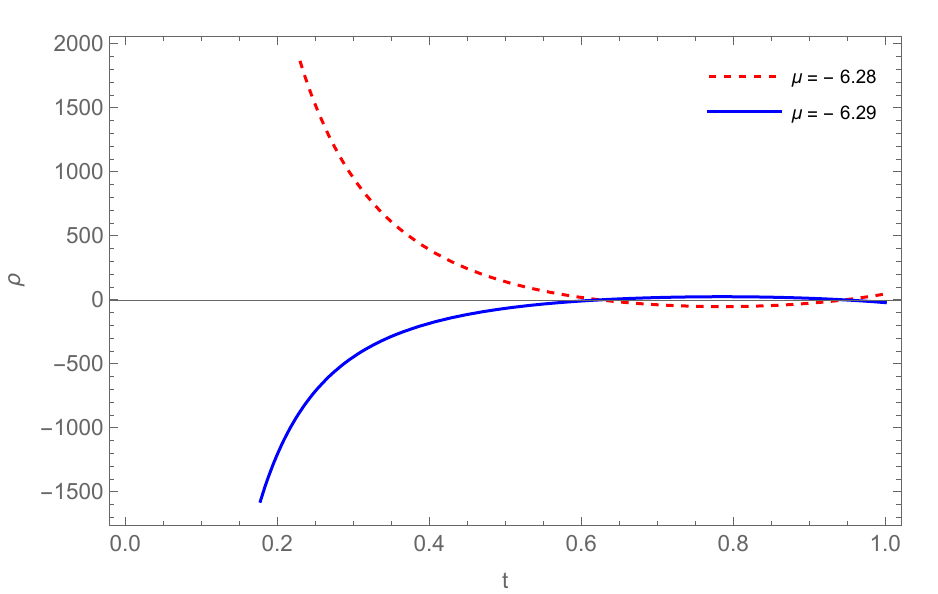}
	\caption{Plot of energy density $ \rho $ \textit{vs.} cosmic time $t$ for $\alpha = k=0.1$}
	\label{fig:1}
\end{figure}
	From figure \ref{fig:1}, it is clear that the energy density is negative for $ \mu=-6.29 $ and positive for $ \mu=-6.28 $, therefore we have a viable model for $\mu\geq -6.28$ with increasing time.

\begin{figure}[htb]
	\centering
	\includegraphics[width=0.7\linewidth]{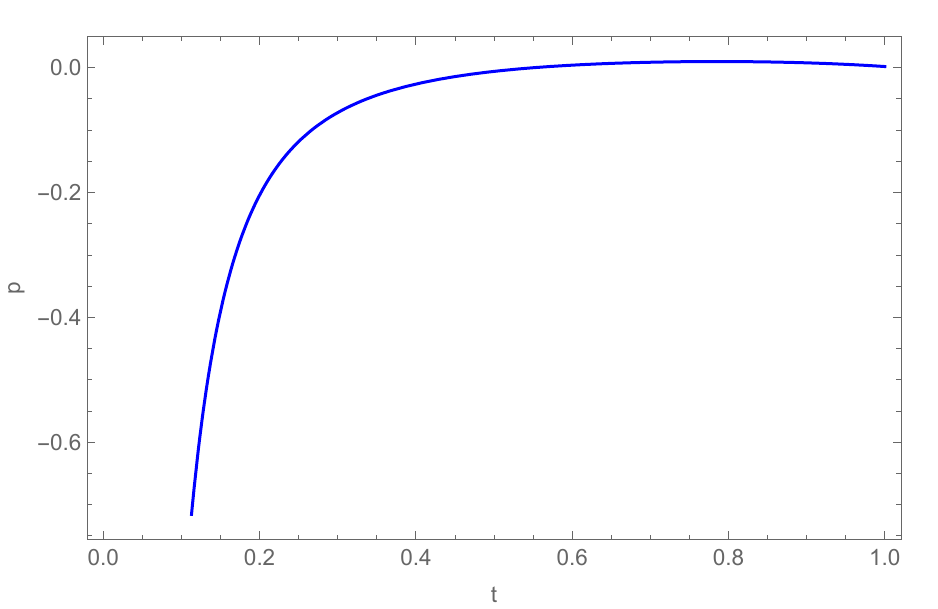}
	\caption{Plot of pressure $ p $ \textit{vs.} cosmic time $ t $ for $\alpha= \mu= k=0.1$}
	\label{fig:2}
\end{figure}

	The pressure of domain wall is negative throughout the evolution of the universe depicted in figure \ref{fig:2}. Thus, domain walls play a role in dark energy. Also, the energy density of barotropic fluid is found to be
\begin{equation} \label{Eq:26}
	\rho_b= \frac{(4\pi+\mu) b^2cos^2(\frac{2kt}{\alpha})}{l_{1}(w+1) sin^2(\frac{2kt}{\alpha})}+{\frac{(-2\pi+3\mu) \alpha^2}{3l_{1}(w+1)k^2}}
\end{equation}
\vspace{-0.5cm}
\begin{figure}[htb]
	\centering
	\includegraphics[width=0.7\linewidth]{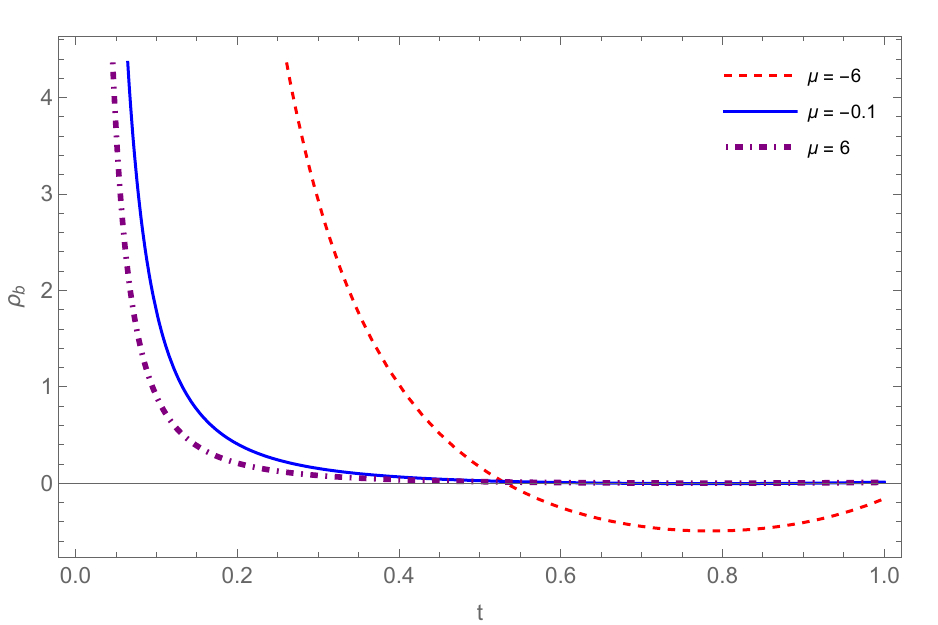}
	\caption{Plot of energy density of barotropic fluid $ \rho_b $ \textit{vs.} cosmic time $ t $ for $\alpha = k=0.1$}
	\label{fig:3}
\end{figure}
	Figure \ref{fig:3} depicts the behaviour of barotropic fluid energy density $ \rho_b $ for different values of parameter $\mu$. It is clear that, for small values of $\mu$,  $ \rho_b $ is decreasing function of $ t $. It is positive and large near $ t=0 $ and tends to zero with increasing time $ t $.
	
	Furthermore, the tension of domain walls $\sigma_d$ is obtained as
\begin{equation} \label{Eq:27}
	\sigma_d= \frac{ (6\pi w+2\pi-\mu) b^2 cos^2(\frac{2kt}{\alpha})}{l_{1}(w+1) sin^2(\frac{2kt}{\alpha})}-{\frac{(8\pi w+6\pi+3\mu) \alpha^2}{3l_{1}(w+1)k^2}}
\end{equation}

\begin{figure}[htb]
	\centering
	\includegraphics[width=0.7\linewidth]{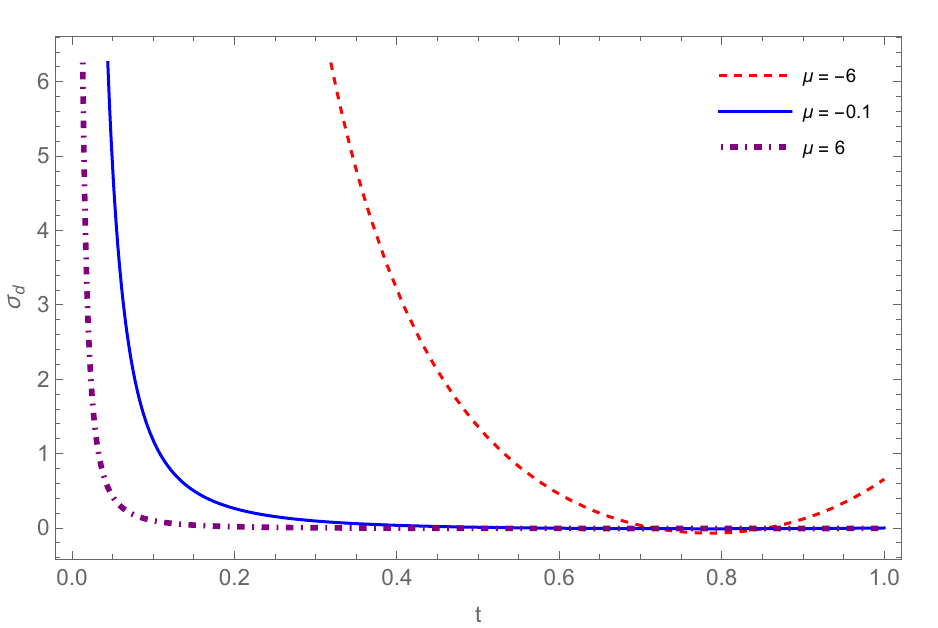}
	\caption{Plot of the tension of domain walls $\sigma_d$ \textit{vs.} cosmic time $ t $ for $\alpha = k=0.1$}
	\label{fig:4}
\end{figure}

	The figure \ref{fig:4} shows the graphical behaviour of the tension of domain walls $\sigma_d$ for $\mu =-6,$ $\mu=-0.1$ and $\mu =6$. It is observed that the tension of domain walls is positive decreasing function of time. Initially, it is very large and decays to zero with increasing time. The surface tension of domain walls control its motion. Zeldovich et al. \cite{ZelDovich et al. 1975} have discussed the properties of domain interfaces, cosmological expansion, and homogeneity of the universe. Domain walls must be disappear at a quite early stage of the evolution of the universe for isotropy and homogeneity. When domain walls disappear, the energy density transformed into energy of massive quanta or into equilibrium radiation. Here,  $\sigma_d > 0$ near $t=0$, which means that domain walls exists in the early stage of the universe. They survive for a very small period of time and vanish, which is as per requirement stated by Zeldovich et al. \cite{ZelDovich et al. 1975}. This confirms our earlier work in $f(R,T)$ theory for FRW and axially symmetric space time \cite{Katore et al. 2016}.

	The physical parameters such as expansion scalar $\theta$, deceleration parameter $q$ and shear scalar $\sigma$ are obtained as
\begin{equation} \label{Eq:28}
	\theta= \frac{2k cos(\frac{2kt}{\alpha})}{\alpha sin(\frac{2kt}{\alpha})} 
\end{equation}
\begin{equation} \label{Eq:29}
	q= -1+3 sec^2(\frac{2kt}{\alpha})
\end{equation}
\begin{equation} \label{Eq:30}
	\sigma= -\frac{kcos(\frac{2kt}{\alpha})}{\sqrt{3}\alpha sin(\frac{2kt}{\alpha})}
\end{equation}
	The sign of deceleration parameter indicates whether the universe is accelerating or decelerating. The positive sign of $ q $ corresponds to decelerating universe, whereas the negative sign corresponds to an accelerating universe. Here, the deceleration parameter is positive i.e. the model indicates a decelerating universe. The expansion scalar $\theta$ and shear scalar $\sigma$ are decreasing function of time; at $t=0$, $\theta$ and $\sigma$ are infinite i.e. the rate of expansion of the universe was very high at the Big Bang. Also they are cyclic. It is noted that $\frac{\sigma}{\theta} \neq0$ and $\frac{\sigma}{\theta}$ is constant, not depends on $t$, therefore the universe is anisotropic throughout the evolution.

	The volume of universe $V$, is found to be
\begin{equation} \label{Eq:31}
	V=\frac{\beta k^2c_1}{\alpha}sin(\frac{2kt}{\alpha})
\end{equation}
	From equation \eqref{Eq:31}, it is clear that the universe is cyclic which undergoes an endless sequence of cosmic epochs that begin with a bang and end in a crunch \cite{Steinhardt and Turok 2002}. Basu \cite{Basu 2008} stated that when the galaxies are found to lie along the curve $q=1$ or above, the universe will expand to a maximum extension and then retrace back to its path in the contraction. The universe contracts to a state of high temperature and density, and at the end of contraction, it will vanish into singularity. The present observational status would suggest that the distance galaxies rather lie along the curve for $q=1$ i.e., the universe is expanding and decelerating.\\
	Liu et al. \cite{Liu et al. 2008} have studied the parametrized equation of the state of dark energy for a variety of models with identical parameters using SNIa data to determine the best way to parametrize it. The parametrize equation of state $ \omega $ in term of average scale factor $ a $ is defined as
\begin{equation}\label{Eq:32}
	\omega=\omega_0+\omega_1(1-\frac{a}{a_0})^r
\end{equation}
	where $\omega_0$, $\omega_1$ are two undefined parameter, $a$ is average scale factor, $a_0$ is the present value of scale factor and $r=1,2... $  Therefore, the new parametrize $\omega$ is obtained as
\begin{equation}\label{Eq:33}
	\omega=\omega_0+\omega_1\left[1-\frac{\beta k^\frac{2}{3}c_1^\frac{1}{3}}{\alpha^\frac{1}{3}}\left(sin(\frac{2kt}{\alpha})\right)^\frac{1}{3}\right]^r
\end{equation}

\begin{figure}[htb]
	\centering
	\includegraphics[width=0.7\linewidth]{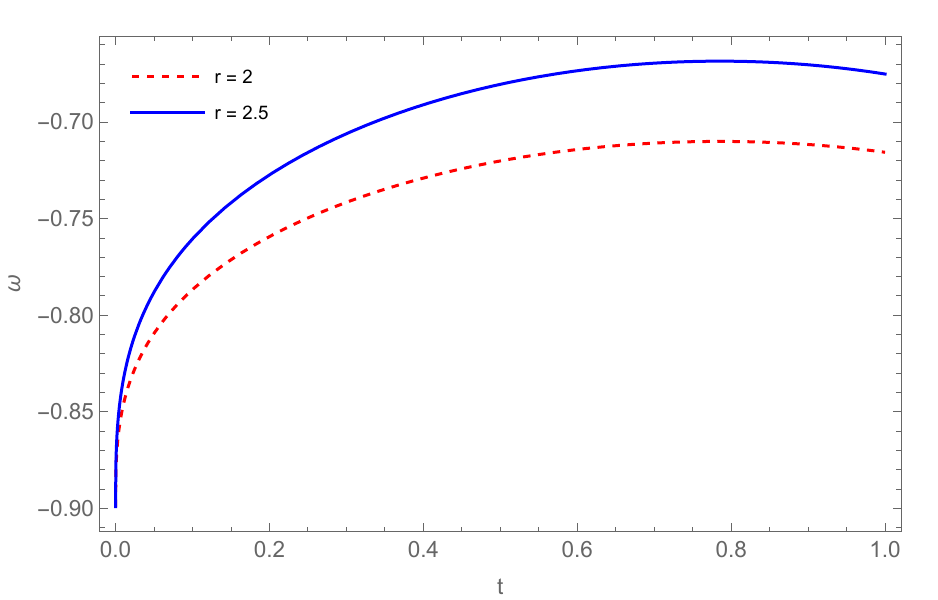}
	\caption{Plot of parametrize $\omega $ \textit{vs.} cosmic time $ t $ for $\alpha = \beta = c_{1} = k=0.1$}
	\label{fig:5}
\end{figure}

	The graphical representation of parametrized $w$ is shown in figure \ref{fig:5}. It is observed that the range of $w$ is $-0.90 \le w \le -0.65$ for the particular values of the parameters $w_{0} = 0.1$ and $w_{1} =-1$. The observational data indicates the range of the EoS parameter $w$ is $-1.013^{+0.068}_{-0.073}$ \cite{Suzuki et al. 2012}, whereas Liu et al. \cite{Liu et al. 2008} obtained the range of the EoS parameter as $-1.5 \le w \le 0.5$. Also, Scolnic et al. \cite{Scolnic et al. 2018} obtained the range of the EoS parameter as $w=-1.026 \pm 0.041$. It is clearly found that the range of EoS parameters obtained in our model is consistent with values obtained by Suzuki et al. \cite{Suzuki et al. 2012} and Scolnic et al. \cite{Scolnic et al. 2018}.

\section{Case II}
	In this case, we consider that the constant of integration is zero i.e. $C_1=0$, then the  equation \eqref{Eq:18} reduced to the following form
\begin{equation}\label{Eq:34}
	L^2B^{2(n+1)}=\frac{\alpha^2}{(n-1)k^2}B^4
\end{equation}
	Simplifying the above equation and integrating, we arrive at
\begin{equation}\label{Eq:35}
	B=E^{\frac{1}{n}}t^{\frac{1}{n}}
\end{equation}
\begin{equation}\label{Eq:36}
	C= \beta E^{\frac{1}{n}}t^{\frac{1}{n}}
\end{equation}
\begin{equation}\label{Eq:37}
	A= kEt
\end{equation}
	where  $E=\frac{n \alpha}{\sqrt{n-1}k}$. It is clear that metric potentials are real for $n>1$ and imaginary for $n<1$.\\
	The expression of energy density and pressure are calculated as follows
\begin{equation}\label{Eq:38}
	\rho= \frac{n+2}{3\mu n^2t^2}+ \frac{(8\pi+\mu)[(2-n)-l_{2}(n+2)]}{l_{1} n^2t^2}
\end{equation}
	where $l_{2}=\frac{8\pi+5\mu}{3\mu}$.

\begin{figure}[htb]
	\centering
	\includegraphics[width=0.7\linewidth]{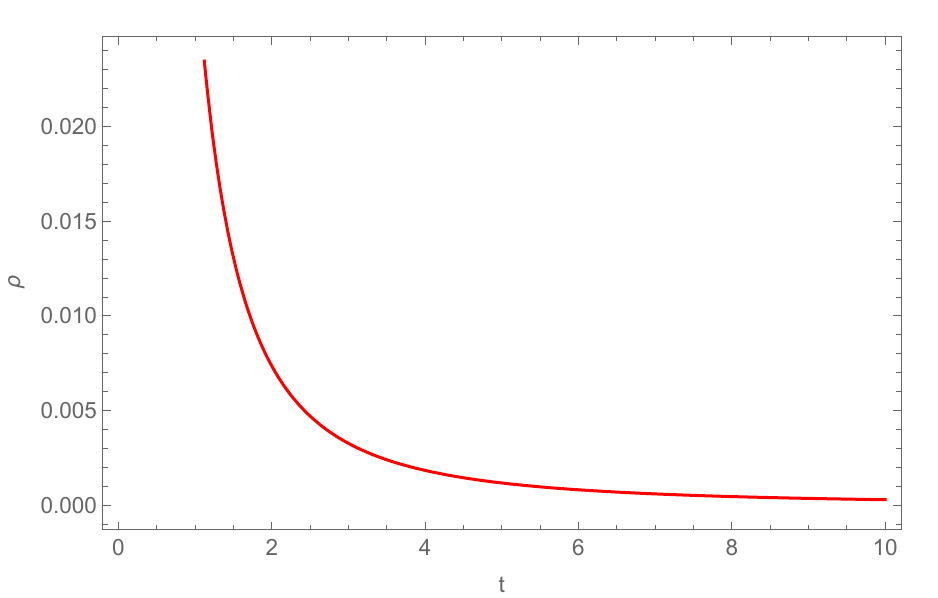}
	\caption{Plot of energy density $\rho$ \textit{vs.} cosmic time $t$ for $\alpha = \mu = 0.1$ and $n=1.1$}
	\label{fig:6}
\end{figure}

	From figure \ref{fig:6}, It is clear that the energy density is a decreasing function of time. It is large near $ t=0 $. A dark energy model for domain walls is also obtained by Caglar and Aygun \cite{Caglar and Aygun 2016} in case of a higher dimensional FRW universe.
\begin{equation}\label{Eq:39}
	p=\frac{3\mu[(2-n)-l_{2}(n+2)]}{l_{1} n^2t^2}
\end{equation}

\begin{figure}[htb]
	\centering
	\includegraphics[width=0.7\linewidth]{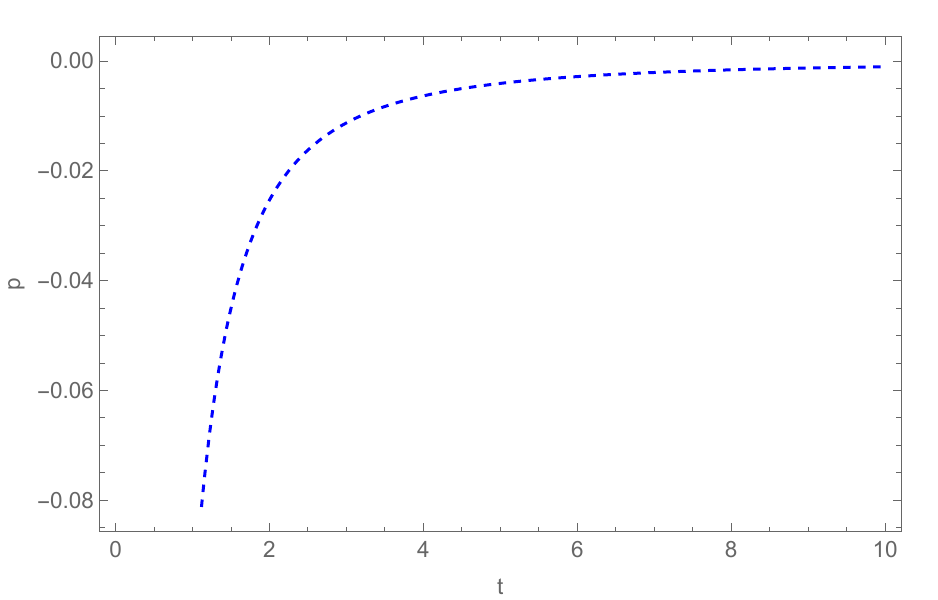}
	\caption{Plot of pressure $p$ \textit{vs.} cosmic time $t$ for $\alpha = \mu = 0.1$ and $n=1.1$}
	\label{fig:7}
\end{figure}

	Also, figure \ref{fig:7} shows that the pressure is negative. Therefore, it is important to note that, we may say, the domain walls acts as dark energy candidates. The energy density of barotropic fluid is found to be
\begin{equation}\label{Eq:40}
	\rho_b= \frac{(2-n)-l_{2}(n+2)}{(8\pi+2\mu)(\omega+1) n^2t^2}+\frac{n+2}{3\mu (\omega+1) n^2t^2}
\end{equation}
\begin{figure}[htb]
	\centering
	\includegraphics[width=0.7\linewidth]{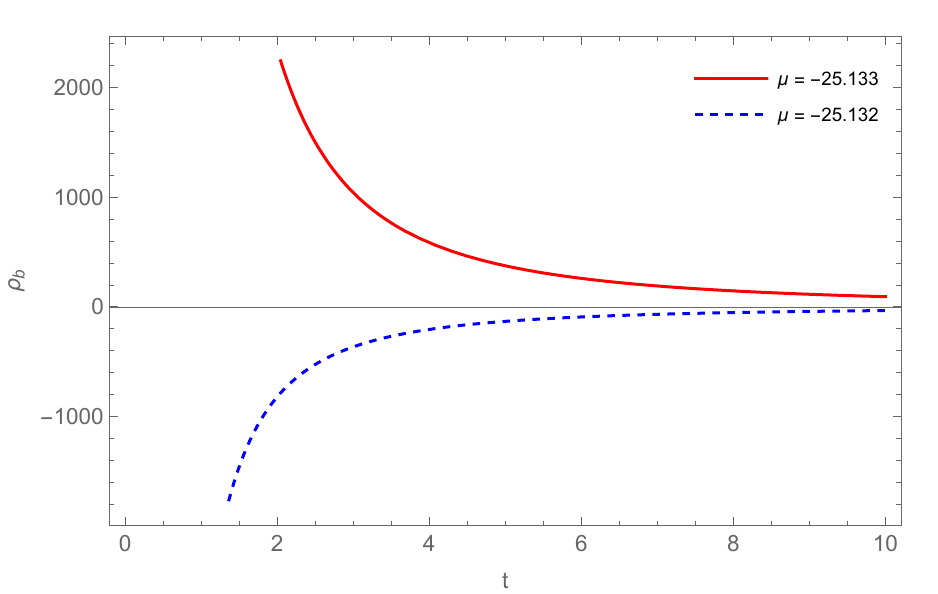}
	\caption{Plot of energy density of barotropic fluid $\rho_b$ \textit{vs.} cosmic time $t$ for $\alpha = 0.1$ and $n=1.1$}
	\label{fig:8}
\end{figure}

	Figure \ref{fig:8} shows that the plot of energy density of a barotropic fluid $\rho_b$ is positive for $n=1.1$, $\mu=-25.133$ and it is negative for $n=1.1$, $\mu=-25.132$. We have a viable model for $\mu\leq-25.133$ and for other values of $\mu$, the energy density is negative.\\
	The tension of the domain walls is obtained as
\begin{equation}\label{Eq:41}
	\sigma_d= \frac{l_{3}[(2-n)-l_{2}(n+2)]}{l_{1} (\omega+1) n^2t^2}+\frac{(n+2) \omega}{3\mu (\omega+1) n^2t^2}
\end{equation}
	where $l_{3}=[(8\pi +\mu) \omega-3\mu]$.
	
\begin{figure}[htb]
	\centering
	\includegraphics[width=0.7\linewidth]{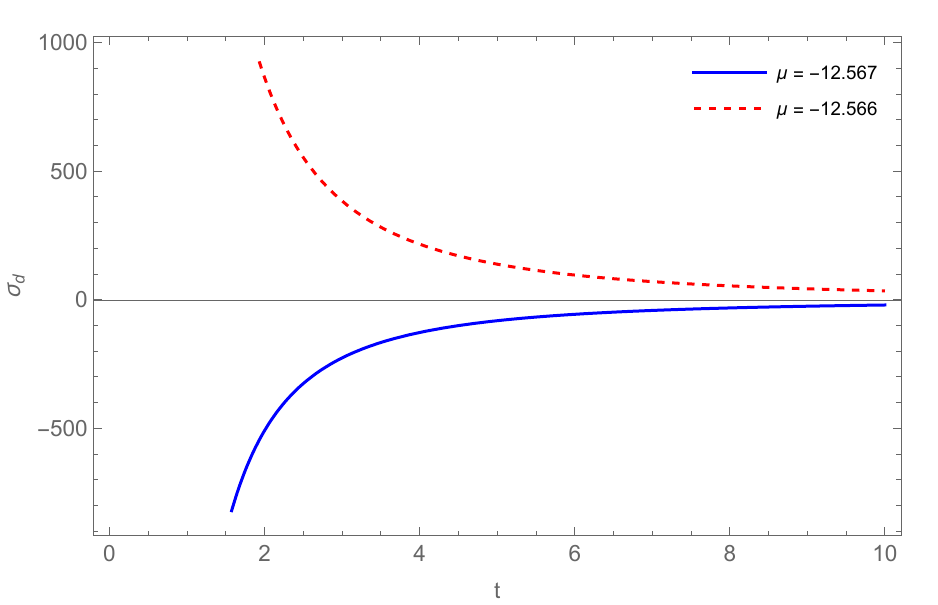}
	\caption{Plot of tension of the domain walls $\sigma_d$ \textit{vs.} cosmic time $t$ for $\alpha = 0.1$ and $n=1.1$}
	\label{fig:9}
\end{figure}

	Figure \ref{fig:9} shows that the graph of tension of domain walls $\sigma_d$ is positive for $n=1.1$ and $\mu=-12.566$, but it is negative for $n=1.1$ and $\mu=-12.567$. The tension of domain walls is negative for a viable model, that is, for $\mu = -12.567$. It indicates that the domain wall transforms into another form of matter. The expansion scalar $\theta$, deceleration parameter $q$ and shear scalar $\sigma$ is obtained as
\begin{equation}\label{Eq:42}
	\theta= \frac{n+2}{nt}
\end{equation}
\begin{equation}\label{Eq:43}
	q= -1+\frac{3n}{n+2}
\end{equation}
\begin{equation}\label{Eq:44}
	\sigma= \frac{n-1}{\sqrt{3}nt}
\end{equation}
	From expression \eqref{Eq:44}, it is clear that the expansion rate of universe is decreasing with increasing time. It is highest near t=0. The shear scalar is also a decreasing function of time. At $n=1$, it is zero. Therefore, the model has isotropy at $n=1$ throughout the evolution. Also, $\frac{\sigma}{\theta}\neq0$ for other values of $n$, $\frac{\sigma}{\theta} \neq0$ i.e. the universe does not approach to isotropy at late time. Recently, Aluri et al. \cite{Aluri et al. 2022} studied consistency of isotropy with observational data and found that there is a deviation from isotropy in topological anomalies in CMB. In the present study, our model does not approach isotropy, and for consistency with the conclusion drawn by Aluri et al. \cite{Aluri et al. 2022}, we will study this in the future work. The deceleration parameter is negative for $n<1$ and positive for $n>1$. As mentioned above, metric potentials $A, B, C$ are real for $n>1$. Therefore, $q$  is positive in that case i.e. the universe is decelerating. The volume of universe $V$ is found to be
\begin{equation}\label{Eq:45}
	V=\beta k\left[\frac{n\alpha t}{k\sqrt{n-1}}\right]^\frac{n+2}{n}
\end{equation}

\begin{figure}[htb]
	\centering
	\includegraphics[width=0.7\linewidth]{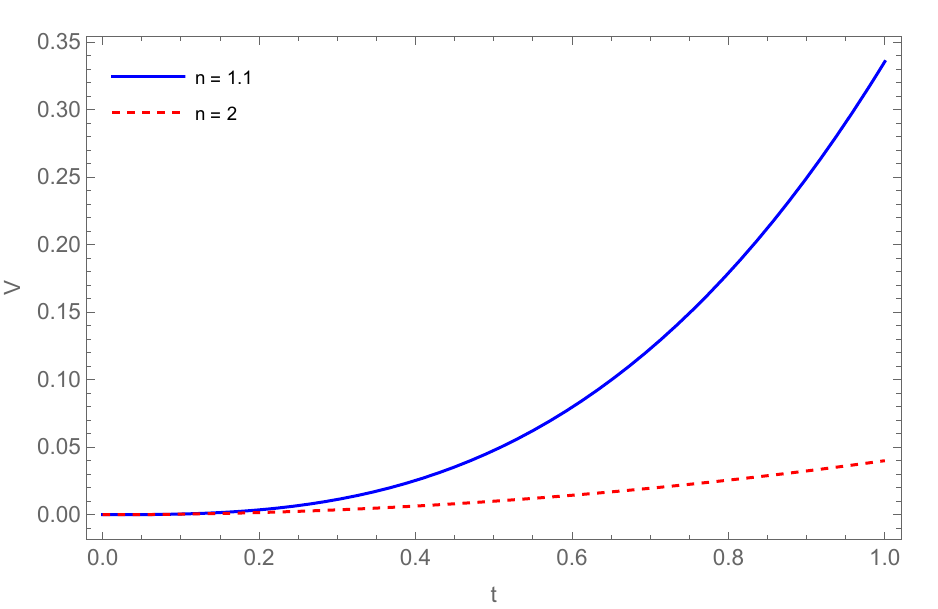}
	\caption{Plot of volume of universe $V$  \textit{vs.} cosmic time $t$ for $\alpha = \beta = k = 0.1$}
	\label{fig:10}
\end{figure}

	From figure \ref{fig:10}, it is clear that the volume of universe is expanding with increasing time. The expansion of the universe is faster for $n=1.1$ than $n=2$. Also, the parametrize equation of state $\omega$ is obtained as
\begin{equation}\label{Eq:46}
	\omega=\omega_0+\omega_1\left[1-\beta^\frac{1}{3}k^\frac{n-6}{3n} \left(\frac{n\alpha t}{\sqrt{n-1}}\right)^\frac{n+2}{n}\right]^r
\end{equation}

\begin{figure}[htb]
	\centering
	\includegraphics[width=0.7\linewidth]{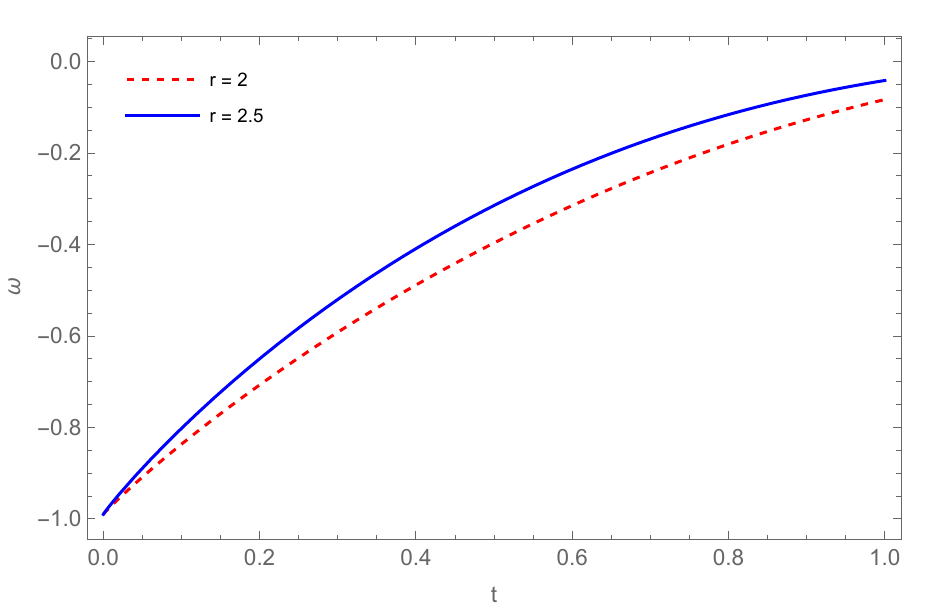}
	\caption{Plot of parametrize $\omega$ \textit{vs.} cosmic time $t$ for $\alpha = \beta = k = 0.1$ and $n=1.1 $}
	\label{fig:11}
\end{figure}
	Figure \ref{fig:11} shows that the graphical representation of parametrized $w$ is an increasing function of cosmic time. For the particular values of the parameters $w_{0} = 0.01$ and $w_{1} =-1$, the range of the EoS parameter is $-1.0 \le w \le -0.1$, i.e., initially the model behaves like $\Lambda$CDM, and as time increases, it behaves like a quintessence dark energy model. In this case, the range of w is smaller than that obtained by Liu et al. \cite{Liu et al. 2008}, Suzuki et al. \cite{Suzuki et al. 2012},  and Scolnic et al. \cite{Scolnic et al. 2018}.
	
\section{Conclusion}
	In this paper, we have studied Bianchi type-$VI_0$  cosmological models in the presence of domain walls in $f(R,T)$ theory of gravity using the function $f(R,T)=R+2f(T)$. The solutions of field equations are determined for the two cases such as $C_1\neq0$, $n=0$ and $C_1=0$.
\begin{itemize}
	\item In case I, the total energy density is positive for $\mu \geq-6.28$ with increasing time. The energy density of barotropic fluid is decreasing as a function of time. It is very large near $t=0$ and diminishes as $t$ tends to infinity. As the tension of domain walls indicates the existence of domain wall, $\sigma_d > 0$, therefore domain walls exists in the early era of the universe and vanish with increasing time by transforming their energy into other forms of matter. In our case, $q$ is positive, which indicates that the decelerating nature of the universe. Moreover, the expansion scalar and shear scalar are periodic functions and their ratio $\frac{\sigma}{\theta}\neq0$, indicating that the universe is anisotropic. Similar results are also obtained by Adhav et al. \cite{Adhav et al. 2009} for domain walls. The novelty of work is that domain walls may behave like dark energy due to the negative pressure. The nature of the universe is cyclic, which begin with Big Bang and end in a Big Crunch.
\end{itemize}
\begin{itemize}
	\item In case II, we observe that the sign of deceleration parameter is negative for $n<1$, thus the universe is accelerating for $n<1$. Also, $q$ is positive for $n>1$ and $q=0$ for $n=1$. As the metric potentials $A,B,C$ are imaginary for $n<1$ and infinite for $n=1$, we discard these values of $n$ and we consider $n>1$, for which we get $q$ positive. Therefore, the universe is decelerating. The total energy density is a decreasing function of time. It is large near $t=0$. We observe that the energy density of barotropic pressure $\rho_b$ is negative, whereas tension of the domain walls $\sigma_d$ is positive. The ratio $\frac{\sigma}{\theta}\neq0$ i.e. the study indicates an anisotropic universe. The universe is expanding.
\end{itemize}

\begin{itemize}
	\item The range of EoS parameters in cases I and II are $-0.90 \le w \le -0.65$ and $-1.0 \le w \le -0.1$, respectively, which resemble the findings of Suzuki et al. \cite{Suzuki et al. 2012} and Scolnic et al. \cite{Scolnic et al. 2018}.
\end{itemize}

\end{document}